\documentclass[aps,prb,reprint,superscriptaddress]{revtex4-2}
\usepackage[colorlinks=true, citecolor=blue, urlcolor=blue ]{hyperref}
\usepackage{epsfig,newlfont,amssymb,amsfonts,amsmath,bm,subfigure,palatino,mathtools,amsthm,braket,soul,enumitem,graphics,graphicx,times,physics, multirow, makecell}
\usepackage[normalem]{ulem}
\usepackage{tabularx}
\usepackage[table,xcdraw]{xcolor}

\begin{document}
\title{
Genuine multipartite entanglement as a  probe of many-body localization in disordered spin chains with Dzya{\l}oshinskii–Moriya interactions
}

\author{Triyas Sapui}
% \email{triyassapui@hri.res.in}
\affiliation{Harish-Chandra Research Institute, Chhatnag Road, Jhunsi, Prayagraj 211019, India}
\affiliation{Homi Bhabha National Institute, Training School Complex, Anushakti Nagar, Mumbai 400094, India}
\author{Keshav Das Agarwal}
% \email{keshavdasagarwal@hri.res.in}
\affiliation{Harish-Chandra Research Institute, Chhatnag Road, Jhunsi, Allahabad - 211019, India}
\affiliation{Homi Bhabha National Institute, Training School Complex, Anushakti Nagar, Mumbai 400094, India}
\author{Tanoy Kanti Konar}
% \email{tanoy.kanti.konar@uj.edu.pl}
\affiliation{Harish-Chandra Research Institute, Chhatnag Road, Jhunsi, Allahabad - 211019, India}
\affiliation{Homi Bhabha National Institute, Training School Complex, Anushakti Nagar, Mumbai 400094, India}
\affiliation{Instytut Fizyki Teoretycznej, Wydzia\l{} Fizyki, Astronomii i Informatyki Stosowanej, Uniwersytet Jagiello\'nski, \L{}ojasiewicza 11, PL-30-348 Krak\'ow, Poland}	
\author{Leela Ganesh Chandra Lakkaraju}
\email[Corresponding author: ]{ganesh.lakkaraju@unitn.it}
\affiliation{Pitaevskii BEC Center, CNR-INO and Dipartimento di Fisica, Universit\`a di Trento, Via Sommarive 14, Trento, I-38123, Italy }
\affiliation{INFN-TIFPA, Trento Institute for Fundamental Physics and Applications, Trento, Italy} 
\author{Aditi Sen (De)}
% \email{aditi@hri.res.in}
\affiliation{Harish-Chandra Research Institute, Chhatnag Road, Jhunsi, Allahabad - 211019, India}
\affiliation{Homi Bhabha National Institute, Training School Complex, Anushakti Nagar, Mumbai 400094, India}

\begin{abstract}
We study the genuine multipartite entanglement (GME) in a disordered system and its potential for detecting the transition from the thermalized to the localized phase. We demonstrate that the quenched average GME can approach its near-maximum value in the ergodic phase of a disordered quantum spin model. In contrast, GME vanishes in the many-body localized (MBL) phase, both in equilibrium and in the long-time dynamical regime, indicating lack of multipartite entanglement in the localized regime. To establish this, we analyze the disordered Heisenberg spin chain subjected to a random magnetic field along with the two- and three-body Dzya{\l}oshinskii–Moriya (DM) interactions. We exhibit that the behavior of GME, in both the mid-spectrum of the Hamiltonian and the dynamically evolved states from an initial N{\'e}el configuration,  serves as a reliable indicator of the critical disorder strength required for the ergodic to MBL transition. The identified transition point aligns well with standard indicators such as the gap ratio and the correlation length. Moreover, we find that the presence of DM interactions, particularly the three-body interaction, significantly enlarges the thermal phase and delays the onset of localization with an extended correlation length. This shift in the transition point is consistently reflected in both static and dynamical analyses, reinforcing GME as a robust probe for MBL transitions.
\end{abstract}
\maketitle

\section{Introduction}

Quantum statistical mechanics posits that isolated, interacting many-body systems generically thermalize under unitary evolution, meaning that local observables reach thermal equilibrium values described by the Gibbs ensemble, effectively erasing memory of initial conditions. This behavior is encapsulated in the eigenstate thermalization hypothesis (ETH)~\cite{Deutsch1991, Srednicki1994, Deutsch_eth}, which asserts that individual energy eigenstates exhibit thermal properties, rendering the system ergodic. ETH explains how unitary evolution aligns with statistical mechanics, as eigenstates in the bulk of the spectrum become locally indistinguishable and featureless. These systems exhibit quantum chaos and ergodicity \cite{DAlessio2016}, reflected in their energy level statistics \cite{Mehta1991, Haake2010, Atas2013} described by the Wigner-Dyson distribution~\cite{Wigner_1951,dyson_1962} in Gaussian random matrix ensembles such as Gaussian orthogonal ensembles (GOE) for real matrices or Gaussian unitary ensembles (GUE) for complex matrices. However, ETH can be violated in systems exhibiting integrability \cite{Rigol2008}, many-body scars \cite{Turner2018, Chandran2023}, Stark localization \cite{Schulz2019}, Floquet dynamics \cite{Ponte2015, Else2016}, or non-Hermitian effects \cite{Cipolloni2024, Roccati2024}, leading to non-ergodic behavior and preventing thermalization.

ETH breaking was first explored through disorder-induced localization in non-interacting systems, known as Anderson localization \cite{Anderson1958}. In recent decades, its many-body counterpart, many-body localization (MBL), emerged in interacting disordered systems
\cite{Oganesyan_huse, Arijeet_huse}. Extensive research has been devoted to this exotic phenomenon across various systems, including spinless fermions~\cite{Oganesyan_huse, Bera2017}, spin chains~\cite{Arijeet_huse, jonas_prl_2014} and the Hubbard model~\cite{Mondaini2015, Sierant2017, bosons_sierant} along with long-range interactions~\cite{Yao2014, Burin2015, Li2016, Nandkishore2017, Tikhonov2018, Sierant2019}, highlighting the apparent universality~\cite{khemani_prl_2017, Zhang2018} and emergent integrability in MBL, characterized by an extensive set of local integrals of motion (l-bits)~\cite{Serbyn2013, Phenomenology_lbit, Imbrie2016}. Despite these efforts and several experimental works~\cite{Smith2016, Kondov2015, Schreiber2015, Choi2016, Rispoli2019, Kohlert2019, Guo2020, Gong2021}, a comprehensive understanding of the transition between the ergodic and MBL phases remains elusive, with several theoretical models proposed ~\cite{ALET2018498, Nandkishore2015, DeRoeck2017, Abanin_review, Sierant2022, Sierant2025}. The key signatures of the MBL phase include Poisson level statistics~\cite{Atas2013, Serbyn2016}, correlation patterns~\cite{Luis2019, Colbois2024}, and quantum order ~\cite{Bauer2013, Huse2013, Chandran2014, Bahri2015, Prakash2017, Friedman2018}. Such systems also possess unique dynamical traits, depicted by the absence of transport~\cite{Agarwal2015}, long-time memory of the initial state(s) characterized by imbalance~\cite{coherence_dynamics, Luitz2016, Doggen2018, Chanda2022}, dynamical ergotropy~\cite{ergotropy_dynmics}, slow growth of entanglement entropy ~\cite{bardasrson_prl_2012, slow_growth_entanglement, Geraedts2017} and magic~\cite{falcao_prl_2025} following a disorder quench. 

The block entanglement entropy, which changes from a volume law in the ergodic phase to an area law in the MBL phase~\cite{Bauer2013, Singh2016, Corps2021, mobility_edge, Anza2020, Aramthottil2021,conc_dynamics, Bera2016} turn out to be key indicators for determining the ergodic-to-MBL transition. Dynamically, disorder-induced slow growth of entanglement also marks the MBL transition~\cite{bardasrson_prl_2012, slow_growth_entanglement, Geraedts2017}. However, studies of multipartite entanglement, a natural faithful quantity for understanding phenomena in many-body systems, remain limited (see recent works considering multisite physical quantities such as quantum Fisher information ~\cite{DeTomasi2019, SafaviNaini2019, Sajna2020, perciavalle2021}, entanglement spectra~\cite{Yang2015, Serbyn2016b}, Schmidt gaps~\cite{Schmidt_gap} and quantum mutual information~\cite{bera_mutualinfo, negativity_bose} although these quantities detect but cannot properly quantify multipartite quantum correlations).

In this work, we address this gap by utilizing a computable measure of genuine multipartite entanglement (GME) for pure states, namely the generalized geometric measure (GGM) ~\cite{Bennett2000, Eisert2001, Verstraete2003, Wei2003, Ma2011, Hashemi2012}, defined via the minimum distance from non-genuinely multipartite entangled states \cite{Sen2010}, to analyze a class of disordered many-body Hamiltonians. Specifically, we investigate a spin-1/2 Heisenberg chain with a randomly chosen magnetic field in the presence of multi-body Dzya{\l}oshinskii–Moriya (DM) interaction~\cite{DZYALOSHINSKY1958241, moriya_1960, moriya_prl_1960,Adani2024}. Originating from spin-orbit coupling, the DM interaction introduces an antisymmetric exchange term between spins and has been widely studied in systems such as the nearest-neighbor and long-range transverse Ising models~\cite{jafri_prb_2008, GR_phase_DM, Konar2025}, the $XY$ model~\cite{jafari_XY_entanglement, roy_prb_2019, Zhong_2013}, and the gamma model~\cite{quantum_phase_DM_2}, where it can induce gapless chiral phases and enhance $zz$-correlations when its strength surpasses anisotropy. Further, it has been found in various solid-state systems, including \( \text{Cu(C}_6\text{D}_5\text{COO})_2\cdot 3\text{D}_2\text{O} \),\( \text{Yb}_4\text{As}_3 \), to name a few ~\cite{comp_1, comp_2,comp_3,comp_4,jafari_experimental}. Moreover, the DM interaction plays a compelling role in the context of many-body localization, as it breaks time-reversal symmetry and changes the spectral statistics from GOE to GUE while preserving the system’s $\mathbb{U}(1)$ symmetry.

We explore the influence of two- and three-body DM interactions in both the ergodic and the MBL regimes. Our analysis reveals that three-body DM interactions have a more significant impact in delaying the ergodic-to-MBL transition than the two-body DM term (cf. ~\cite{Stagraczynski2017, Cao2023}), thereby extending the thermal phase. The three-body DM interaction considered here is conventionally known as the $XZY-YZX$-type three-site interaction in quantum spin chains and has been investigated in connection with chiral phases, asymmetric quasiparticle spectra, magnetoelectric responses, and nontrivial dynamical quantum phase transitions~\cite{You2016,Cao2022,Thakur2023}. We analyze the eigenstates of the disordered Hamiltonian and find that the profile of quenched average genuine multipartite entanglement, measured by GGM~\cite{Sen2010}, successfully tracks the ergodic to MBL transition, which is in good agreement with conventional measures such as the spectral statistics and the long-range correlator. In particular, the quenched average GGM values and the most probable value of GGM are high in the ergodic phase but drop significantly in the MBL phase.

By initializing the system in a Néel state and evolving it under the same class of Hamiltonians, we demonstrate that the trends of multipartite entanglement during the long-time regime reveal a delayed ergodic-to-MBL transition in the presence of DM interactions, marked by distinct changes in the proposed scaling laws. These observations align well with known indicators such as imbalance~\cite{Luitz2016, Doggen2018, Chanda2022}. Notably, high multipartite entanglement is generated in both short and long time periods of the ergodic phase, whereas the MBL phase exhibits significantly reduced entanglement generation. This comprehensive study highlights the impact of DM interactions on localization phenomena and firmly establishes multipartite entanglement as a reliable indicator for disorder-induced phase transitions.

The paper is organized as follows. In Sec. \ref{sec:model_methods}, we describe the model and the relevant quantifiers such as GGM in Sub-Sec. \ref{sec:methods}, level statistics in Sec. \ref{sec:level_statistics_definition}, and long-range classical correlators in Sec. \ref{sec:long_range_correlator_definition}. We then use the GGM to characterize the MBL to ergodic transition and compare it with the traditional measures in Sec. \ref{sec:equilibrium_study}, which is further divided into the study of the Heisernberg model without and with DM interaction in Secs. \ref{sec:Heisenberg} and \ref{sec:DM_dependency}, respectively. We also perform finite-size scaling in Sec. \ref{sec:finite_size_scaling}. The dynamics of GGM is studied in Sec. \ref{sec:dynamics}, and we finally conclude in Sec. \ref{sec:conclusion}.

\section{Introducing model and transition indicators}
\label{sec:model_methods}

We consider a spin-$1/2$ one-dimensional Heisenberg model with two- and three-body Dzya{\l}oshinskii–Moriya interactions~\cite{DZYALOSHINSKY1958241, moriya_1960, moriya_prl_1960}, subject to a random external magnetic field along the $z$-direction, which introduces disorder into the system. The Hamiltonian reads as
\begin{align}
H &= J \sum_{k=1}^{N}  \vec{S}_k \cdot \vec{S}_{k+1}
+ \tilde{D} \left( S_k^x S_{k+1}^y - S_k^y S_{k+1}^x \right) \nonumber \\
&+ 2\tilde{D^\prime}  \left( S_k^x S_{k+1}^z S_{k+2}^y - S_k^y S_{k+1}^z S_{k+2}^x \right) + \tilde{h}_k S_k^z,
\label{eq:hamil}
\end{align}
where $S^a_k = \sigma^a_k/2$ $(a = \{x,y,z\})$ are the spin operators and $\sigma^a_k$ are the Pauli matrices at site $k$ with periodic boundary conditions ($S^a_{k+N}\equiv S^a_k$). To eliminate energy dependence in all figures of merit, we redefine the random magnetic field as $h_k = \tilde{h}_k/J$, sampled uniformly from the interval $[-h, h]$ and \( D = \tilde{D}/J \) and \( D^\prime = \tilde{D}^\prime/J \), denoting the strengths of the two-body and three-body DM interactions, respectively. The $\mathbb{U}(1)$ symmetry present in the disordered Heisenberg chain is preserved by the inclusion of both two- and three-body DM interactions, as they conserve the total magnetization, $S^z = \sum_{k=1}^{N} S_k^z$, i.e., $[H, S^z] = 0$ even when $D, D^\prime \neq 0$. This allows the Hamiltonian to be block-diagonalized into magnetization sectors labeled by $S^z \in \{-N/2, -N/2+1, \dots, N/2\}$. In our analysis, we restrict ourselves to the $S^z = 0$ sector (and even $N$), which has the largest Hilbert space among all $S^z$ blocks. Without the DM interactions, i.e., \(D=D'=0\), the model is known as the disordered Heisenberg model and is known to undergo a transition from the ergodic to the MBL phase~\cite{Arijeet_huse} with increasing disorder strength. 

Let us briefly distinguish the two types of DM interactions appearing in Eq.~(\ref{eq:hamil}) (the second and third terms). The term proportional to $D$ is the conventional
nearest-neighbor DM interaction, $D(\boldsymbol{S}_k\times\boldsymbol{S}_{k+1})_z$, describing a bilinear antisymmetric exchange between two neighboring spins \cite{DZYALOSHINSKY1958241,moriya_1960}. On the
other hand, the coupling constant $D'$ is conventionally referred to as an $XZY-YZX$-type three-body interaction and can be written as $2D'\sum_kS_{k+1}^{z}\left(\boldsymbol{S}_{k}\times\boldsymbol{S}_{k+2}\right)_z$.
Throughout this work, we refer to this term as the three-body DM interaction, owing to its antisymmetric DM-like structure. It couples the spins at sites $k$ and $k+2$ through the $z$ component of the intermediate spin and is therefore distinct from the conventional DM interaction both in its multisite structure and in its effective interaction range. Moreover, such $XZY-YZX$-type interactions are known to generate chiral phases, asymmetric quasiparticle spectra, and nontrivial magnetoelectric and dynamical properties in quantum spin chains \cite{You2016,Cao2022,Thakur2023}.

In this work, we focus on the effects of two- and three-body DM interactions on the transition and utilize an information-theoretic quantity for its detection, which also provides information about the model's suitability in quantum information processing tasks.

{\it Set the stage to identify the transition point.} To characterize the phases in this model, we choose a physical quantity of interest and perform quenched averaging over different disorder realizations. In particular, the given quantity is computed by averaging over both the eigenstates and the disorder realizations. We consider the $n_{\varepsilon}$ number of eigenstates in the middle of the spectrum of the Hamiltonian to compute the average of a quantity $O$ as $\langle O\rangle=\frac{1}{n_{\varepsilon}}\sum_{j=1}^{n_{\varepsilon}}O (| \psi_j\rangle)$ where \(|\psi_j\rangle\) belongs to the $n_{\varepsilon}$ eigenstates with eigenvalue close to $E_{\text{mid}}=(E_{{\max}}+E_{{\min}}){/2}$ (with $E_{{\max(\min)}}$ corresponding to the maximum(minimum) eigenvalue). Subsequently, an averaging over $n_R$ randomly chosen disorder realizations is carried out to compute the quenched average quantity as $\overline{\langle O \rangle}$. The eigenstates are obtained via exact diagonalization~\cite{arma2025} till $N=14$, and POLFED~\cite{Sierant2020} technique is used for system-sizes \(N = 16, 18\) and $20$ (see Appendix~\ref{app:polfed}). We select the corresponding values of $n_{\varepsilon}$ and $n_R$ by performing a convergence check up to the desired accuracy for a given system-size $N$  (see Table \ref{tab:eigenstate_label}), by computing the standard deviation of averaged values of $n_R/10$ realizations. In the case of dynamics, a disorder average over $n_R^\prime$ realizations is taken to compute the quenched average dynamical quantity, $\overline{O(t)}=\frac{1}{n_R^\prime}\sum_{p=1}^{n_R^\prime}O (| \psi_{p}(t)\rangle)$, of time evolved states $\ket{\psi_p(t)}$, starting from the N{\'e}el state. (see Table \ref{tab:eigenstate_label}). The behavior of $\overline{O(t)}$ is used to distinguish between the ergodic and the MBL phases.

\begin{table}
    \centering
    {\setlength{\tabcolsep}{0pt}
    \begin{tabularx}{0.95\linewidth}{|
        >{\columncolor{gray!20}\centering\arraybackslash}X|
        >{\centering\arraybackslash}X|
        >{\centering\arraybackslash}X|
        >{\centering\arraybackslash}X|
        >{\centering\arraybackslash}X|
        >{\centering\arraybackslash}X|
        >{\centering\arraybackslash}X|
        >{\centering\arraybackslash}X|}
        \Xhline{2\arrayrulewidth}
    \textbf{$N$} 
      & \(8\) & \(10\) & \(12\) & \(14\) & \(16\) & \(18\) & \(20\) \\
      \Xhline{2\arrayrulewidth}
    \textbf{$n_{\varepsilon}$}
      & \(22\) & \(80\) & \(200\) & \(400\) & \(800\) & \(1000\) & \(1000\) \\
      \Xhline{2\arrayrulewidth}
    \textbf{$n_R$}
      & \(10^4\) & \(10^4\) & \(5000\) & \(2000\) & \(500\) & \(100\) & \(100\) \\
      \Xhline{2\arrayrulewidth}
    \textbf{$n_R^\prime$}
      & \(5000\) & \(3000\) & \(2000\) & \(1000\) & \(1000\) & \(300\) & \(-\) \\
      \Xhline{2\arrayrulewidth}
    \end{tabularx}}
    \caption{The table indicates the number of eigenstates $n_{\varepsilon}$ picked from the middle of the spectrum and  $n_R$ is the number of disorder realizations used to perform quenched averaging for a given system size $N$. \(n_R'\) indicates the number of disorder realizations taken in the case of dynamics while performing disorder averaging.}
    \label{tab:eigenstate_label}
\end{table}

\subsection{Multipartite entanglement quantifier to reveal transition}
\label{sec:methods}

Let us first introduce the information-theoretic measure used to characterize MBL in this work, in particular, the genuine multipartite entanglement measure, which is a distance-based measure in the state space~\cite{ggm_shimony}.

% \subsubsection{Quantifier for genuine multiparty entanglement}
% \label{sec:ggm_definition}
A pure \(N\)-party state $| \Psi_{1,2,\ldots,N}\rangle$, is genuinely multiparty entangled (GME) if it is not product across any bipartition. Based on Fubini-Study metric, quantification of GME content of a pure state is possible, known as generalized geometric measure (GGM)~\cite{Wei2003, Sen2010, ggm_shimony,Barnum_ggm_2}. It is defined as the minimum distance from the set of all non-genuinely (nG) multipartite entangled states, i.e.,
\begin{equation}
    \mathcal{G}(|\Psi\rangle)=1-\underset{|\chi\rangle}{\max}|\langle\chi|\Psi\rangle|^2,
\end{equation}
where \(|\chi\rangle\) belongs to the set of nG entangled states. An equivalent expression for the above can be given as
\begin{eqnarray}
    \mathcal{G}(| \Psi\rangle)=1-\max_{A:B}[\lambda_{A:B}^2|A\cap B=\emptyset,|A\cup B|=N],
\end{eqnarray}
in which the maximization is carried out over all bipartitions \(A:B\), and \(\lambda_{A:B}\) denotes the largest Schmidt coefficient in the corresponding bipartition and it is easy to verify that \(\lambda_{A:B}^2\) is an eigenvalue of the local density matrix. Although it is a computable measure for moderate \(N\) in arbitrary dimensions, the computation of \(\mathcal{G}(\ket{\Psi})\) becomes numerically intensive because of the exponential increase of the number of such bipartitions with the system-size \(N\). Hence, this computational complexity necessitates the use of approximations in calculating the GGM.
 \begin{figure}  
    \centering \includegraphics[width=\linewidth]{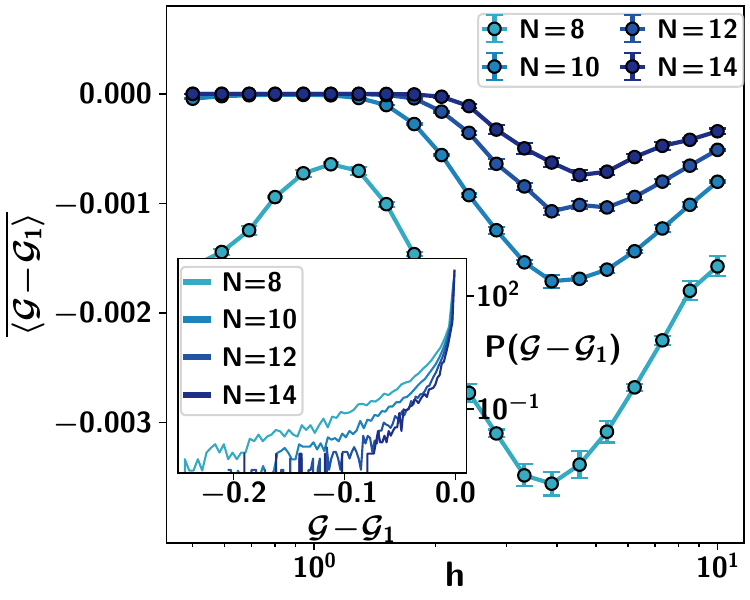}
    \caption{The difference between the exact GGM, $\mathcal{G}$, and its single-site approximation, \(\mathcal{G}_1\), i.e., \(\overline{\langle \mathcal{G} - \mathcal{G}_1\rangle}\) (ordinate)  against the disorder strength, \(h\) for increasing system-sizes of the Heisenberg model, i.e., $D=D^\prime=0$. It  reveals that as the system-size grows, the difference vanishes. This trend suggests that \(\overline{\langle \mathcal{G}_1\rangle}\) serves as an accurate alternative for the genuine multiparty entanglement  and can be used in place of \(\overline{\langle \mathcal{G}\rangle}\) in these systems. (Inset) Probability distribution \(P(\mathcal{G}\!-\!\mathcal{G}_1)\) of the difference  is plotted as a function of the difference \(\mathcal{G}-\mathcal{G}_1\) for a disorder strength \(h=4.55\) where the deviation \(\overline{\langle \mathcal{G}-\mathcal{G}_1\rangle}\) is maximum. Here $P(X)$ denotes the fraction of the total number of states with $|X(\ket{\psi_j})-X|<5\times 10^{-4}$. Large deviations of $\mathcal{G}$ from $\mathcal{G}_1$ are exponentially suppressed with large system-sizes resulting in error bars $\lesssim 10^{-4}$ for $N=8$ and $\lesssim 10^{-5}$ for higher system-sizes. All axes are dimensionless.}
    \label{fig:ggm_vs_gm}
\end{figure}

To mitigate this challenge, we simplify our approach by restricting attention to Schmidt coefficients obtained from all single-site reduced density matrices. With this simplification, we define the geometric measure (GM), also a multipartite entanglement measure, as
\begin{align}
    \mathcal{G}_1(| \Psi\rangle)=1-\max_{A:B,|A|=1}[\lambda_{A:B}^2|A\cap B=\emptyset,|A\cup B|=N],
\end{align}
where the maximization is performed only over bipartitions for which \(A\) consists of a single spin. To validate the effectiveness of this approximation in the problem addressed here, we compute both $\mathcal{G}$ and $\mathcal{G}_1$ for all eigenstates of various system-sizes. We find that for our Hamiltonian, \(|\overline{\langle \mathcal{G} - \mathcal{G}_1\rangle}|\approx e^{-N}\) (see Fig. \ref{fig:ggm_vs_gm}) which decreases rapidly with increasing system-size, indicating that \(\mathcal{G}_1\) is as effective as \(\mathcal{G}\) in characterizing multipartite entanglement for large system sizes. Therefore, in the rest of this paper, we use \(\mathcal{G}\) and \(\mathcal{G}_1\) interchangeably, as there is no significant distinction (up to two decimal places) between these two quantities to quantify the phases which can help in studying the role of two- and three-body DM interaction in altering the phases in Heisenberg spin chain. Note that the computational resource for calculating GGM ($\mathcal{G}_1$) of a generic eigenstate of a many-body Hamiltonian does not grow significantly with system size as it depends only on the single-site density matrices of the eigenstate. Since $\mathcal{G}_1$ depends only on the largest eigenvalues of the single-site reduced density matrices, it also offers a potential route for experimentally probing multipartite entanglement without requiring full quantum-state tomography. A possible implementation based on randomized measurements \cite{preskill_rand_toolbox} is discussed in  Appendix~\ref{app:randomised_measurement_toolbox}.

\begin{figure}  
    \centering 
    \includegraphics[width=\linewidth]{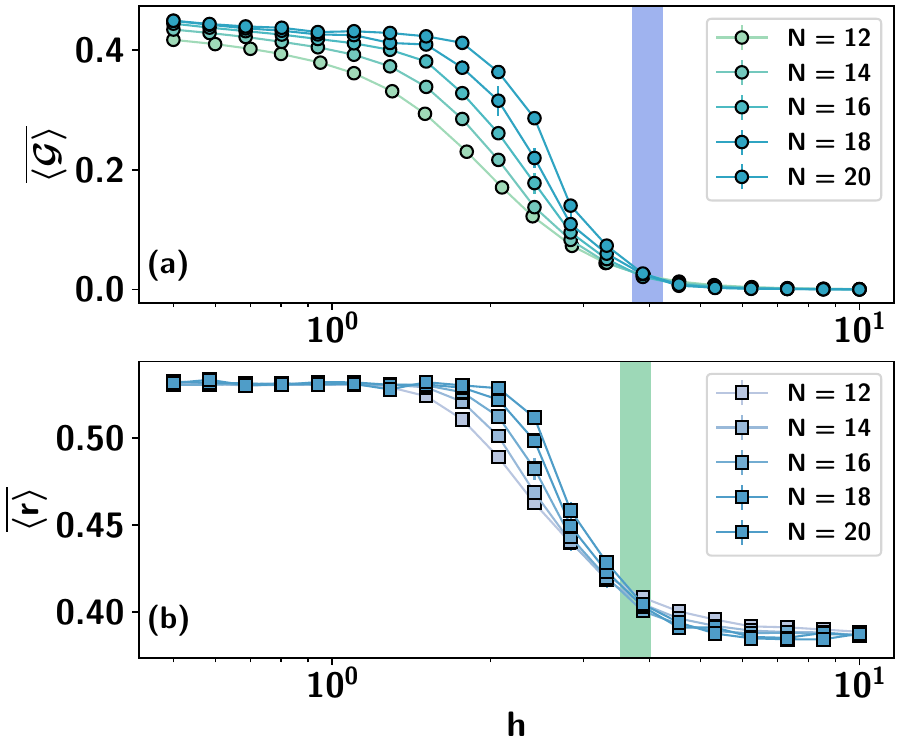}
    \caption{GGM as a detector of MBL in the disordered Heisenberg model. (a) The quenched average GGM, \(\overline{\langle\mathcal{G}\rangle}\) (ordinate), and (b) the average gap ratio, \(\overline{\langle r\rangle}\) (ordinate), as functions of the disorder strength \(h\) (abscissa). Both quantities are plotted for increasing system-sizes \(N\), illustrating that \(\overline{\langle\mathcal{G}\rangle}\) serves as an equally effective indicator as \(\overline{\langle r\rangle}\) in identifying the ergodic to MBL phase transition in the Heisenberg model ($D = D^\prime =0$). The critical disorder strength is estimated to be \(h^* \sim 4.0\) for GGM, while for the gap ratio,  \(h^* \sim 3.8\) is obtained. All axes are dimensionless.}
    \label{fig:A}
 \end{figure}

\subsection{Conventional transition detectors}

We now briefly discuss two physical quantities that are commonly used to probe the ergodic-to-MBL transition.

\subsubsection{Level statistics for benchmarking}
\label{sec:level_statistics_definition}
A method, referred to as level statistics, has been extensively utilized in the literature to identify ergodic and MBL phases because of its distinct behavior in these two phases of the disordered model. Since its inception, it has been established that the energy level spacings of a non-integrable Hamiltonian typically follow a Wigner-Dyson distribution~\cite{Wigner_1951,dyson_1962}, while those of an integrable Hamiltonian adhere to a Poisson distribution~\cite{Atas2013, Serbyn2016}. One commonly used quantity to characterize level statistics is the gap ratio, defined as
\begin{eqnarray}
    r_n = \frac{\min(\varDelta_n, \varDelta_{n-1})}{\max(\varDelta_n, \varDelta_{n-1})},
\end{eqnarray}
where \(\varDelta_n = E_{n+1} - E_n\) represents the spacing between two consecutive energy levels, \(n\) and \(n+1\) with \(\{E_j\}\) representing the eigenenergies of the Hamiltonian. A quenched average gap ratio, \(\overline{\langle r \rangle} = 0.5307\) indicates that the system is ergodic and falls within the Gaussian orthogonal ensemble (GOE), corresponding to the Wigner-Dyson distribution. Conversely, if \(\overline{\langle r \rangle} = 0.3863\), the level statistics is Poissonian, signifying that the system resides in an MBL phase. Moreover, for systems that break time-reversal symmetry, the spectral statistics follow the Gaussian unitary ensemble (GUE), characterized by \(\overline{\langle r \rangle} = 0.5996\).

\begin{figure*}  
    \centering 
    \includegraphics[width=\linewidth]{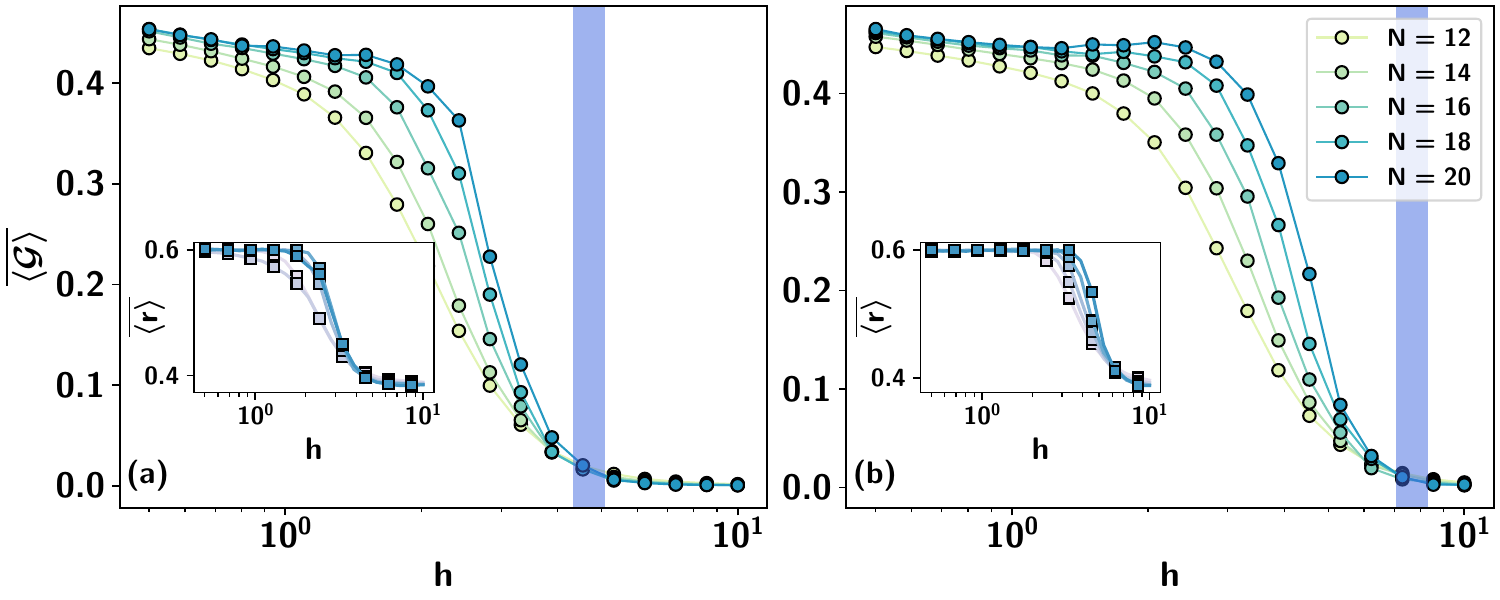}
    \caption{Delayed MBL with DM interactions. Average GGM \(\overline{\langle\mathcal{G}\rangle}\) (ordinate) against disorder strength \(h\) (abscissa) for increasing system-sizes, considering both (a) two-body (\(D=0.5, D^\prime=0\)) and (b) three-body (\(D=0.0, D^\prime=0.5\)) Dzya{\l}oshinskii–Moriya (DM) interactions. These plots exhibit a transition from the ergodic phase to the many-body localized one, similar to the transition identified using the gap ratio \(\overline{\langle r \rangle}\) (shown in the insets of both figures). The critical disorder strength is estimated by the measure \(\overline{\langle\mathcal{G}\rangle}\) to be \(h^* \sim 4.7\) for the two-body DM interaction and \(h^* \sim 7.7\) for the three-body DM interaction, indicated by a solid vertical bar. It is evident that DM interactions can extend the ergodic phase and delay the transition, which is substantial for three-body DM interactions. All axes are dimensionless.
}
    \label{fig:gm_dm}
 \end{figure*}

\subsubsection{Long-range classical correlators}
\label{sec:long_range_correlator_definition}
Another useful transition indicator is the average long-range classical correlator, defined as
\begin{eqnarray}
    C_r^{zz}\equiv C_{ij}^{zz}(\ket{\psi})=\langle \psi|S_{i}^{z}S_{j}^{z}|\psi \rangle -\langle\psi|S_{i}^{z}|\psi\rangle \langle \psi |S_{j}^{z}|\psi\rangle,
    \label{eq:correlator_definition}
\end{eqnarray}
where \(|i-j|=r\) and \(\ket{\psi}\) is a state in the middle of the energy spectrum of the Hamiltonian. Pal and Huse argued that in the ergodic phase, \(\langle \psi |S_{j}^{z}|\psi\rangle \sim 0\) because of the system being in the \(S^z = 0\) sector, although a finite long-range correlation due to the first term persists between distant spins~\cite{Arijeet_huse}. In contrast, in the MBL phase, \(\langle \psi |S_{j}^{z}|\psi\rangle\) acquires a finite value and the long-range correlation decays exponentially with increasing spin separation. These can be analyzed by the behavior of \(\overline{\langle\ln |C^{zz}_r|\rangle}\), which has been shown to detect the transition from the ergodic to the MBL phase~\cite{Arijeet_huse}.

As \(\mathcal{G}\) probes the multipartite entanglement structure of the eigen- (and dynamical) states of the disordered quantum systems, it is interesting to examine how \(\overline{\langle\mathcal{G} \rangle}\) detects the transition point in the presence of two- and three-body DM interactions in comparison with \(\overline{\langle r \rangle}\) and \(\overline{\langle\ln |C^{zz}_r|\rangle}\).

\section{Characterizing phases with  Dzya{\l}oshinskii-Moriya interaction}
\label{sec:equilibrium_study}

To establish the GGM as a faithful quantity to indicate the ergodic to MBL transition, we first study the well-studied Heisenberg model, i.e., when $D= D^\prime = 0$ in Eq. (\ref{eq:hamil}) and then carry out the investigation for $D\neq 0$ as well as  $D^\prime \neq 0$. We then report the performance of GGM to differentiate the phases in the presence of two- and three-body DM interactions. In addition, we propose the distribution of the GGM over the eigenstates of the Hamiltonian as an indicator that can differentiate two phases. 

\subsection{MBL and ergodic phases in Heisenberg model}
\label{sec:Heisenberg}
 
A clear one-to-one correspondence between \(\overline{\langle\mathcal{G}\rangle}\) and \(\overline{\langle r\rangle}\), indicates that multisite entanglement effectively captures the MBL transition [see Fig.~\ref{fig:A}(a) for \(\overline{\langle\mathcal{G}\rangle}\) and Fig.~\ref{fig:A}(b) for \(\overline{\langle r\rangle}\)]. Specifically, for the Heisenberg spin chain model (\(D=D^\prime=0\)), the transition is observed near \(h^* \sim 4.0 \) when the measure is taken to be \(\overline{\langle\mathcal{G}\rangle}\), which is in good agreement with previous results~\cite{Arijeet_huse,mobility_edge}, indicating the MBL transition at \( h^* \sim 3.8\). This transition point can be consistently verified by the standard indicator \(\overline{\langle r\rangle}\), which takes the value \(\overline{\langle r\rangle} \sim 0.53\) in the ergodic phase (\(h < h^*\)) while it drops to \(\overline{\langle r\rangle} \sim 0.39\) in the MBL phase (\(h > h^*\)). On the other hand, in the ergodic regime (\(h < h^*\)), \(\overline{\langle\mathcal{G}\rangle}\) saturates to a finite value, approximately \(\overline{\langle\mathcal{G}\rangle} \lesssim 0.5\), with this feature becoming more pronounced as the system size increases. The value of \(\overline{\langle\mathcal{G}\rangle}\) decreases for moderate \(h\) values and finally vanishes in the MBL phase, as expected. As the GGM quantifies (along with  the detection~\cite{Li2026} of) the genuine multipartite entanglement of pure states, the average GGM $\overline{\langle\mathcal{G}\rangle}$ decreases gradually with increasing disorder strength, signifying a crossover from the ergodic to MBL regime. Significantly, the transition becomes sharper with increasing system size, as presented in Fig.~\ref{fig:A}(a). This result suggests that the MBL phase may be unsuitable for building quantum devices that rely on high multipartite entanglement to function effectively.

Although the suppression of entanglement with increasing disorder is expected, the GGM provides information beyond the entanglement across a fixed bipartition. For example, a state of the form $\ket{\psi}=\ket{\phi}_{1}\otimes\ket{\omega}_{2,\ldots,N}$ has $\mathcal{G}(\ket{\psi})=0$, even when
$\ket{\omega}_{2,\ldots,N}$ can be entangled, and the half-chain von-Neumann entropy can be nonvanishing. Therefore, a vanishing GGM does not necessarily
imply a completely factorized state, but indicates the absence of genuine $N$-partite entanglement. Interestingly, the average entanglement between the nearest-neighbor sites is low in the ergodic phase, and high in the MBL phase~\cite{Bera2016}, in contrast to the average GGM. Moreover, for the finite systems considered here, $\overline{\langle\mathcal{G}\rangle}$ decreases
progressively over a range of disorder strengths rather than vanishing abruptly at a single value of $h$, thereby revealing the gradual suppression of genuine multipartite correlations across the ergodic-to-MBL crossover.

Let us provide an intuitive explanation for why GGM effectively captures the MBL transition, as well as the saturation of \(\overline{\langle\mathcal{G}\rangle}\) to characteristic values in both the ergodic and the MBL regimes. This behavior of \(\overline{\langle\mathcal{G}\rangle}\) can be understood through the lens of random matrix theory (RMT). In the ergodic phase, the system's Hamiltonian is expected to follow RMT predictions, where the eigenstates resemble Haar-random pure states. From the perspective of entanglement theory, Haar-random multipartite states are known to exhibit high multipartite entanglement on average with an increase in the number of sites~\cite{Gross2009, Bremner2009, Rethinasamy2019}. Consequently, the eigenstates in the thermal (ergodic) regime display a high average GGM, close to the maximum value of \(\lesssim 0.5\). On the other hand, as the disorder strength increases and the system enters the MBL phase, multipartite entanglement gets strongly suppressed, leading to vanishing GGM. These findings reinforce the validity of \(\overline{\langle\mathcal{G}\rangle}\) as a robust and insightful marker of the MBL transition, and the above argument suggests its potential universality across disordered systems.

\begin{figure}  
    \centering
    \includegraphics[width=\linewidth]{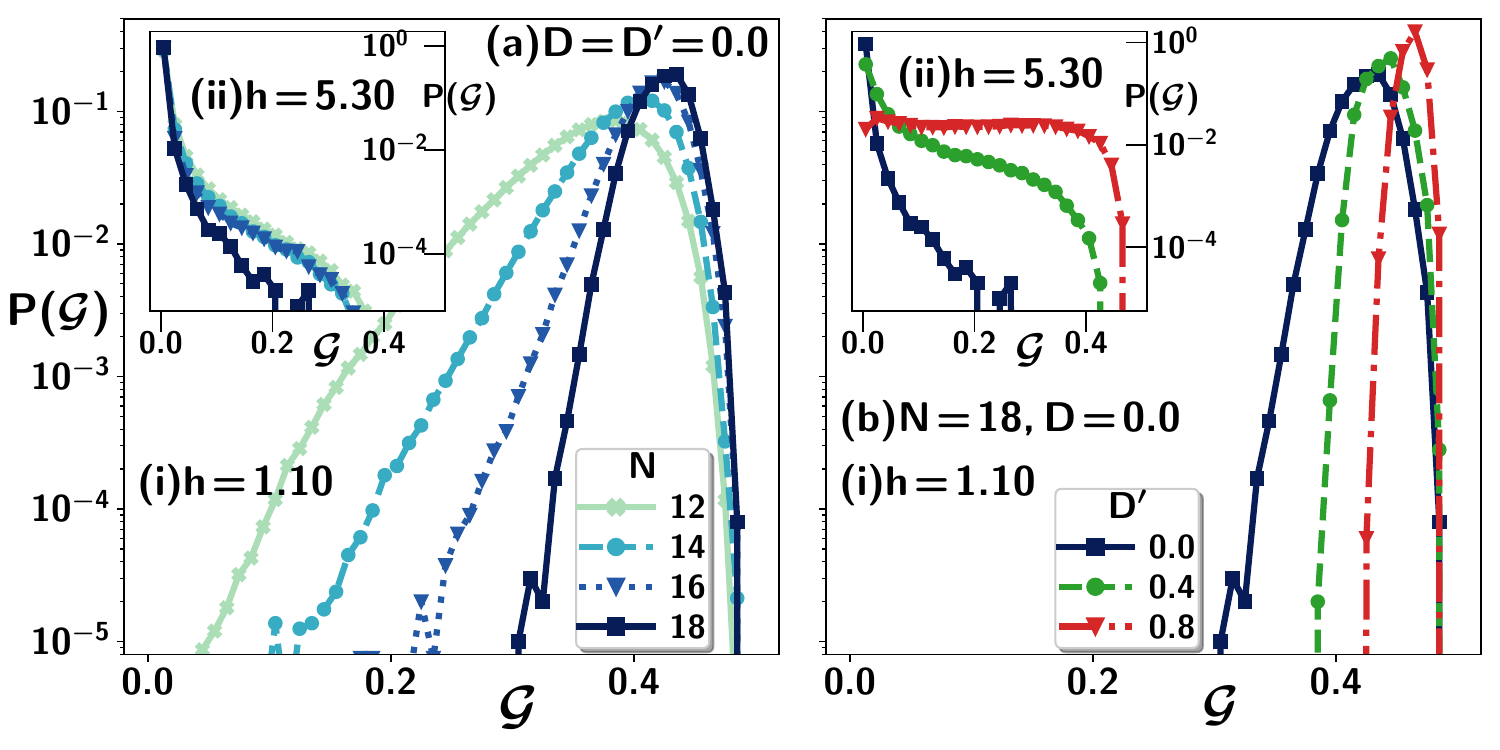}
    \caption{Distribution of GGM $\mathcal{G}(\ket{\psi_i})$, \(P(\mathcal{G})\) (ordinate) vs \(\mathcal{G}\) for eigenvectors $\ket{\psi_i}$ in the middle of the spectrum with various realizations of uniformly distributed magnetic field $h_k\in[-h, h]$ at each site $k$. (a) For the Heisenberg model, i.e., $D=D^\prime=0$, the GGM of states tends to $\mathcal{G}\gtrsim 0.4$ for a large number of states with increasing system-size $N$ in the thermalized regime (shown for $h=1.1$). (Inset of (a)) In the localized region for $h=5.3$, a larger number of states have small GGM, $\mathcal{G}\lesssim 0.05$, on increasing system-size $N$. (b) GGM of the ensemble in the thermalized regime increases with increasing three-body DM interactions $D^\prime$, both in the previously ($D=D^\prime=0$) thermalized and localized [inset of (b)] regions, thereby extending the thermalized regime. All the axes are dimensionless.}

    \label{fig:gm_2d_1.0}
 \end{figure}

\subsection{Extended ergodic phase in Heisenberg model with DM interaction detectable via GGM}
\label{sec:DM_dependency}

The nearest-neighbor DM interaction arises from spin-orbit coupling, induces long-range correlation among distant spins and manifests as an antisymmetric exchange interaction between spins~\cite{DZYALOSHINSKY1958241, moriya_1960, moriya_prl_1960}. Such interactions have recently become experimentally accessible in quantum simulators~\cite{kunimi_pra_2024}. We explore here the impact of the DM interaction in modulating the transition from the ergodic to the many-body localized phase in the Heisenberg chain.

{\it Influence of two-body DM interactions on transition.} The inclusion of a two-body DM interaction alters the behavior of both \(\overline{\langle\mathcal{G}\rangle}\) and the gap ratio. Importantly, since the DM interaction breaks time-reversal symmetry, as mentioned before, the spectral statistics exhibit behavior consistent with the GUE class, instead of GOE. Hence, in the ergodic regime, \(\overline{\langle r \rangle}\) approaches \(0.6\) as the system-size \(N\) increases, which is characteristic of the GUE statistics, as shown in the inset of Fig.~\ref{fig:gm_dm}(a) by the solid lines. The level statistics exhibit a transition from the ergodic to the MBL phase as the disorder strength increases, with \(\overline{\langle r \rangle}\) decreasing from \(0.6\) to \(0.39\). However, in the presence of DM interaction, the ergodic phase is extended, requiring a higher disorder strength to induce the transition to the MBL phase. Interestingly, the trends of \(\overline{\langle\mathcal{G}\rangle}\) also signal a delayed ergodic-to-MBL transition, as stronger disorder is needed for the GGM to completely vanish in the MBL phase. Specifically, we observe that for $D=0.5$ $(D^\prime=0)$, $h^*\sim 4.7$ [see Fig.~\ref{fig:gm_dm}(a)]. Note that the location of the transition is dictated by the strength of the DM interaction, which implies a greater delay in the transition with higher $D$ values.
This observation suggests that the ergodic phase becomes more robust under the influence of DM interactions~\cite{Stagraczynski2017, Cao2023}. In this regime, the enhanced correlations introduced by the DM term necessitate a stronger disorder to localize the system and induce the MBL phase.

{\it Three-body DM interaction delays the transition significantly.} Instead of the two-body DM interaction, let us incorporate three-body DM interactions, i.e., $D=0, D^\prime\ne 0$ in Eq.~(\ref{eq:hamil}). Again, the DM interaction is responsible for further postponement of the transition point, i.e., $h^*$. In particular, the transition to the MBL phase is further delayed compared to the two-body interaction, with a critical disorder strength of approximately \(h^* \sim 7.7\), [see Fig.~\ref{fig:gm_dm}(b)] obtained via \(\overline{\langle\mathcal{G}\rangle}\) which is also in good agreement with $h^*\sim6.2$, predicted by \(\overline{\langle r\rangle}\). These values are calculated via analysis of finite-size scaling in Sec.~\ref{sec:finite_size_scaling}. This prolonged thermal phase can be attributed to the next-nearest-neighbor (NNN) hopping induced by the three-body DM interaction. Such hopping enhances delocalization, thereby promoting ergodicity. However, strong local perturbations, such as a disordered external magnetic field, suppress this hopping mechanism, thereby facilitating localization and leading to the MBL phase at sufficiently high disorder strength. The next step is to investigate the distribution of GGM across the middle spectrum of eigenstates and how this distribution alters with the introduction of the DM interaction. 

\begin{figure}
    \centering
    \includegraphics[width=\linewidth]{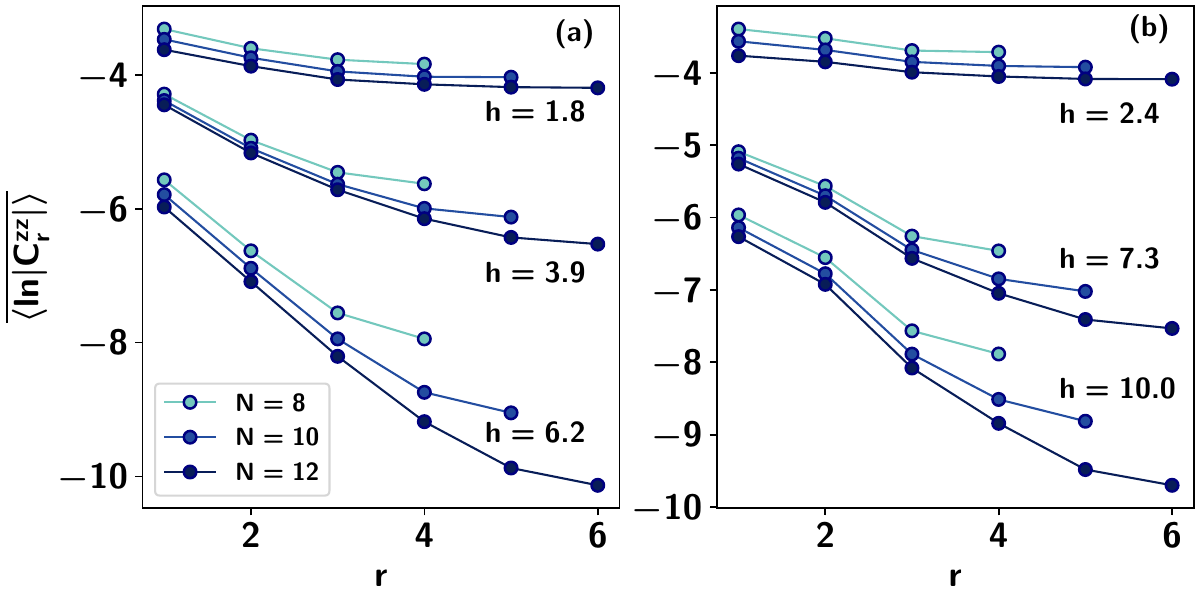}
    \caption{Long-range correlations \(\overline{\langle\ln |C^{zz}_r|\rangle}\) (ordinate) against the distance between the spins, \(r\) (abscissa). We plot the decay of long-range correlation in the \(z\) direction with \(r\) for different disorder strengths with two- and three-body DM interactions for different system sizes. (a) For the two-body DM interaction ($D=0.5, D^\prime=0$), in the ergodic phase \(\overline{\langle\ln |C^{zz}_r|\rangle}\) saturates to a high fixed value, denoting the presence of strong correlations arising from the DM term. On the other hand, in the MBL phase,  such correlation decays exponentially as seen for \(h=6.2\). (b) Effect of the three-body DM interaction—a similar trend is observed for $D=0$, and $D^\prime=0.5$, at higher disorder strength \(h\). For instance, corresponding to \(h=6.2\) in (a), similar behavior emerges when \(h=10\). Therefore, a higher disorder $h$ is required for the transition from the ergodic to the MBL phase. All the axes are dimensionless.}
    \label{fig:czz_dm}
\end{figure}

\emph{Distribution of GGM as a transition marker.} An important question that arises is how multipartite correlations, such as GGM \(\mathcal{G}\) of the eigenstates, are distributed across the ergodic and MBL phases. We demonstrate that in the ergodic phase, the distribution of \(\mathcal{G}\) is sharply peaked near its maximum value, indicating that most eigenstates exhibit nearly maximal multipartite entanglement. The frequency distribution of $\mathcal{G}$ is defined as $P(\mathcal{G})=\frac{|\mathcal{S}_\mu(\mathcal{G})|}{n_R n_\varepsilon}$, where $|\mathcal{S}_\mu(\mathcal{G})|=|\{\ket{\psi_j}, \mathcal{G}(\ket{\psi_j})\in [\mathcal{G}-\mu,\mathcal{G}+\mu]\}|$ is the cardinality of the set of eigenstates with the GGM value $\mathcal{G}(\ket{\psi_j})$ $\mu$-close to $\mathcal{G}$ ($\mu=5\times10^{-3}$ for Fig.~\ref{fig:gm_2d_1.0}). The observations in the absence of DM interactions can be listed as follows: (1a) In the absence of DM interaction, $P(\mathcal{G})$ corresponding to the large \(\mathcal{G}\) value ($\gtrsim0.4$) dominates, while the distribution exhibits an exponential decay as \(\mathcal{G}\) deviates from its maximum. (1b) System size also plays a significant role—as the system size increases, the peak of the \(\mathcal{G}\) distribution converges towards a saturated value ($\gtrsim0.4$). (1c) In the MBL phase, the peak of the distribution shifts toward lower values of \(\mathcal{G}\)($\lesssim 0.05$), closer to zero, indicating that eigenstates in the MBL regime possess significantly lower multipartite entanglement [see the inset of Fig.~\ref{fig:gm_2d_1.0}(a)]. (2a) When two- and three-body DM interactions are introduced, the ergodic phase is extended, and the peak of the \(\mathcal{G}\) distribution quickly shifts towards larger values as the strength of the DM interaction increases [see Fig.~\ref{fig:gm_2d_1.0}(b) for the effect of three-body DM interaction]. In other words, with the increase of $D$ or $D^\prime$, the standard deviation of $P(\mathcal{G})$ decreases. (2b) Additionally, the frequency of states with  large \(\mathcal{G}\) increases with the increasing DM interaction strength, suggesting that DM interactions enhance long-range correlations among spins. Consequently, the number of highly entangled eigenstates grows with increasing DM interaction strength.

\emph{Increment of long-range correlation in presence of DM interaction.} Let us now argue that the delay in the ergodic to MBL transition in the presence of DM interactions may be attributed to persistent LR correlations, \(\overline{\langle\ln |C^{zz}_r|\rangle}\), in the systems. For all values of $h$, it decays with distance $r$, although the decay rate varies with the disorder strength. In particular, when $D=D^\prime=0$, the decay of \(\overline{\langle\ln |C^{zz}_r|\rangle}\) is slower for $h$ belonging to the deep ergodic phase while its decay is rapid in the MBL phase \cite{Arijeet_huse}. In the presence of DM interaction, we find that even for moderate to large $h$ values, the decay rate is small and it gets larger only when the disorder strength is strong enough (see Fig.~\ref{fig:czz_dm}). Further, the LR correlation value itself of the Heisenberg model with non-vanishing DM interaction becomes higher than in the case without DM interactions. For example, at a fixed system size $N=12$ and distance $r=6$, $\overline{ \langle\ln|C^{zz}|\rangle}\sim -10.5$ for $h=6.23$ and $\overline{ \langle\ln|C^{zz}|\rangle}\sim -4.5$ for $h=1.77$ for $D=0.5,D^\prime=0$ [see Fig. \ref{fig:czz_dm}(a)], while similar $\overline{ \langle\ln|C^{zz}|\rangle}$ values ($\sim-10.5$ and $\sim-4.5$) are obtained for higher $h$ ($h=10.0$ and $h=2.41$, respectively), when $D=0,D^\prime=0.5$ [see Fig. \ref{fig:czz_dm}(b)].
As a result, stronger disorder strength is needed to observe the onset of exponential decay, indicating a delayed transition to the MBL phase. In both the two-body and three-body DM interaction scenarios, the long-range correlation decays exponentially at sufficiently large disorder strength, which is a hallmark of the MBL phase. These findings confirm that the DM interaction enhances the correlations between spins and effectively extends the ergodic phase which is also responsible for high multipartite entanglement in the system.

\section{finite-size scaling analysis for transition}
\label{sec:finite_size_scaling}

\begin{figure}
    \centering
    \includegraphics[width=\linewidth]{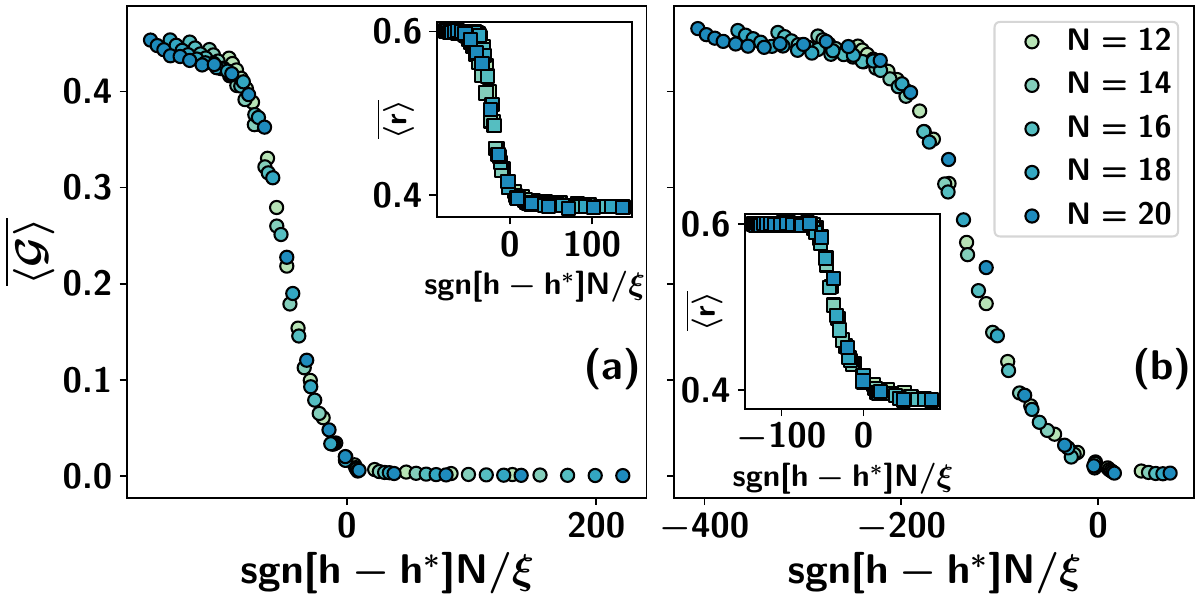}
    \caption{Finite-size scaling analysis of the averaged GGM and level spacing (in insets) in the presence of (a) two- and (b) three-body DM interactions. For five different system-sizes, \(N=12, 14,  16, 18\) and \(20\), \(\overline{\langle G \rangle}\) (ordinate) against \(sgn[h-h^*]N/\xi\) (abscissa). 
    Interestingly, we obtain the scaling exponent and critical disorder strength, which are independent of system size. The resulting values and the exponents obtained are reported in Table \ref{tab:scaling_results}. All the axes are dimensionless.}
    \label{fig:finite_scaling}
\end{figure}
The divergence of the correlation length is a central concept in the theory of critical phenomena and phase transitions. The transition from the ergodic to the many-body localized phase is typically characterized by a power-law divergence of the correlation length of classical correlators. In such scenarios, the correlation length (of finite-size systems) diverges as
\begin{equation}
    \xi = \frac{1}{|h - h^*|^{\nu}},
\end{equation}
where \(\nu\) is the critical exponent. \(h^*\) denotes the finite-size estimate of the critical disorder strength. In this study, we fix the value of \(h^*\) (independent of system-size $N$) and perform a finite-size scaling analysis to obtain scale-invariant observables $\tilde{X} \equiv X(N/\xi)$. This can be accomplished by minimizing a cost function $C^X$ \cite{jonas_prl_2014,mobility_edge,khemani_prl_2017, Zhang2018, mode_prb_2020}, to quantify the quality of data collapse. The cost function is given by

\begin{equation}
    C^{X} = \frac{\sum_{k=1}^{N_{\text{tot}} - 1} \left| X_{k+1} - X_k \right|}{\max\{X_k\} - \min\{X_k\}} - 1,
\end{equation}
where \(N_{\text{tot}}\) is the number of data points corresponding to different values of disorder strength \(h\), system size $N$, and all \(X_k\) values are sorted in non decreasing order according to the rescaled variable \(\text{sgn}[h - h^*]N/\xi\). An ideal data collapse yields a cost function \(C^X = 0\). We perform finite-size scaling analysis for both the average gap ratio \(\overline{\langle r\rangle}\) and the multipartite entanglement measure \(\overline{\langle \mathcal{G}\rangle}\), for the random Heisenberg model as well as for cases involving two- and three-body DM interactions.

\begin{table}[h!]
\centering
{\setlength{\tabcolsep}{0pt}
\begin{tabular}{|>{\centering\arraybackslash}p{0.3\linewidth}|>{\centering\arraybackslash}p{0.2\linewidth}|>{\centering\arraybackslash}p{0.1\linewidth}|>{\centering\arraybackslash}p{0.1\linewidth}|}
\hline
\rowcolor{gray!20} \textbf{Interaction Type} & \textbf{Quantity} & \(\boldsymbol{h^*}\) & \(\boldsymbol{\nu}\) \\
\hline
\multicolumn{1}{|>{\centering\arraybackslash}p{0.3\linewidth}|}{\multirow{2}{*}{\(D=0,~D'=0\)}} 
& \(\overline{\langle r\rangle}\) & 3.76 & 1.33 \\
\cline{2-4}
& \(\overline{\langle\mathcal{G}\rangle}\) & 3.96 & 1.26 \\
\hline
\multicolumn{1}{|>{\centering\arraybackslash}p{0.3\linewidth}|}{\multirow{2}{*}{\(D=0.5,~D'=0\)}} 
& \(\overline{\langle r\rangle}\) & 4.03 & 1.08 \\
\cline{2-4}
& \(\overline{\langle\mathcal{G}\rangle}\) & 4.69 & 1.44 \\
\hline
\multicolumn{1}{|>{\centering\arraybackslash}p{0.3\linewidth}|}{\multirow{2}{*}{\(D=0,~D'=0.5\)}} 
& \(\overline{\langle r\rangle}\) & 6.22 & 1.06 \\
\cline{2-4}
& \(\overline{\langle\mathcal{G}\rangle}\) & 7.66 & 1.53 \\
\hline
\end{tabular}
}
\caption{ Finite-size estimates of the critical disorder strength \(h^*\) and critical exponent \(\nu\) from scaling analysis of \(\overline{\langle r \rangle}\) and \(\overline{\langle \mathcal{G} \rangle}\) for different interaction types.  The \(h^*\) obtained from GGM increases with two- and three-body DM interactions, which is in good agreement with the values from the gap ratio. This also establishes  the usefulness of GGM for the detection of ergodic to MBL phases.}
\label{tab:scaling_results}
\end{table}

\begin{figure}
    \centering
    \includegraphics[width=\linewidth]{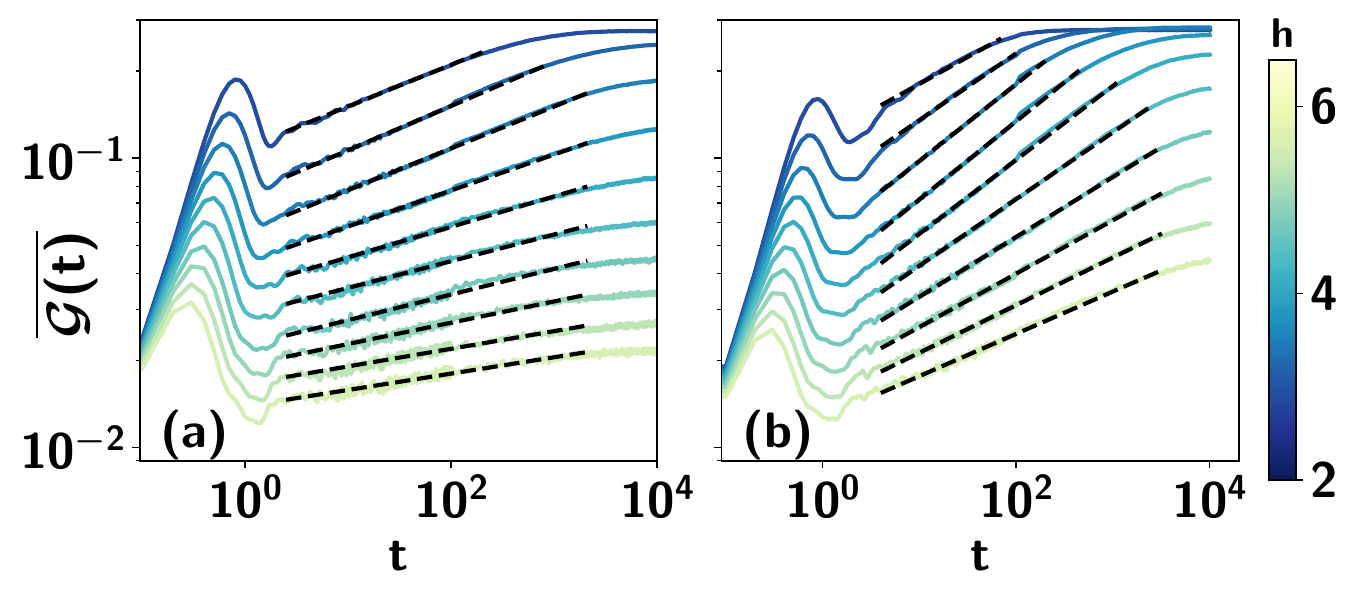}
    \caption{Dynamics of genuine multipartite entanglement. Averaged GGM \(\overline{\mathcal{G}(t)}\) (ordinate) of the time-evolved state, \(\ket{\psi(t)}\), with time $t=J\tilde{t}$ for a disordered Heisenberg spin chain with DM interactions for different disorder strengths [Eq. (\ref{eq:hamil})] with the initial state being the N{\'e}el state, \( |\psi(0)\rangle = |\uparrow \downarrow \uparrow \downarrow \ldots \uparrow \downarrow \rangle \).  (a) The two-body DM interaction with $D = 0.5$ and (b) three-body  non vanishing DM interaction with,  $ D^\prime = 0.5$ and \(D =0\). Different curves correspond to distinct disorder strength values \(h\), ranging from the ergodic regime (\(h = 2.0\)) to many-body localized phases (\(h = 6.5\)), The dashed lines represent the polynomial growth of \(\overline{\mathcal{G}(t)}\) in transient time. Here, the system size $N=18$ and all the axes are dimensionless.}
    \label{fig:ggm_dyn}
\end{figure}

\begin{figure*}
    \centering
    \includegraphics[width=\linewidth]{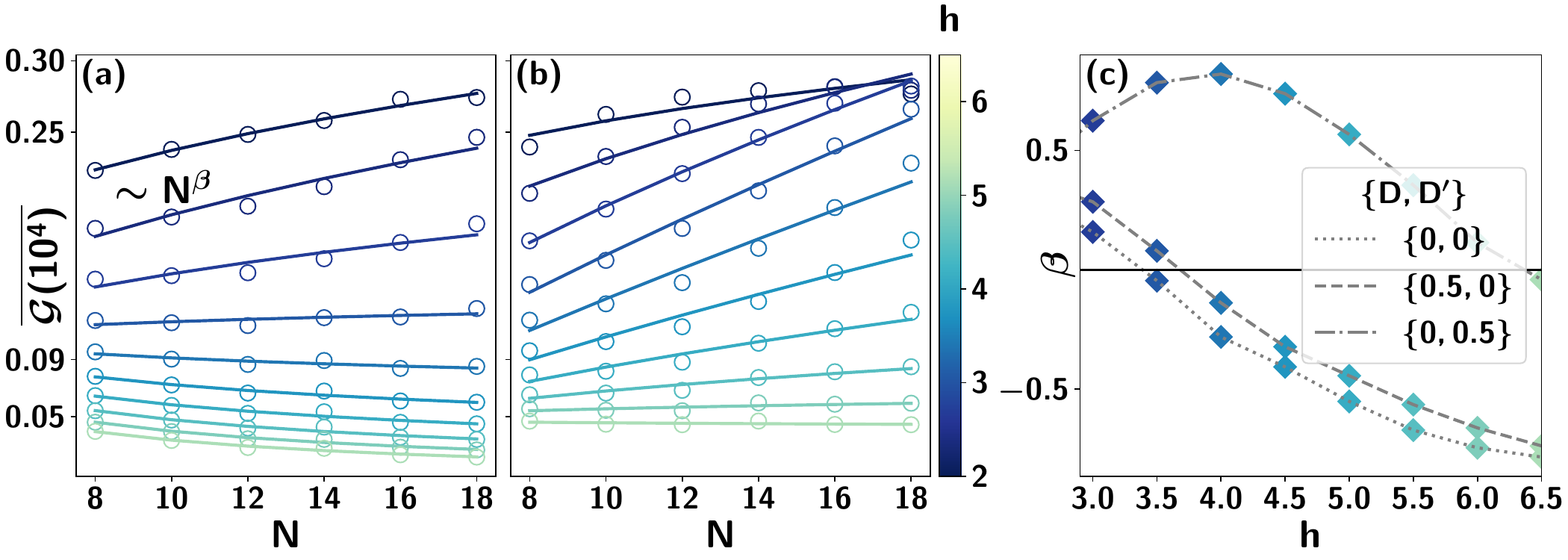}
    \caption{Detection of transition point from the long-time regime. (a) and (b) \(\overline{\mathcal{G}(10^4)}\) (vertical axis) with system-size \(N\) (horizontal axis) for two-body ($D=0.5, D^\prime=0.0$) and three-body ($D=0.0, D^\prime=0.5$) DM interactions respectively. Each curve corresponds to a distinct disorder strength \(h\), ranging from the ergodic regime (\(h = 2.0\)) to many-body localized phases (\(h = 6.5\)), as indicated in the legend. (c) Fitting \(\overline{\mathcal{G}(10^4)} \sim  cN^\beta \), the exponent \(\beta\) (vertical axis) against \(h\) (horizontal axis) for three sets of \(D\) values. The transition point is given by disorder strength $h^*$, where \(\overline{\mathcal{G}(10^4)} \sim  N^0\), i.e., independent of system-size. All the axes are dimensionless.}
    \label{fig:ggm_steady_state}
\end{figure*}

In order to find out system-size-independent criticality in the transition from the ergodic to the MBL phase, we perform finite-size scaling in Fig. \ref{fig:finite_scaling}. The finite-size estimates of the critical disorder strength, \(h^*\) and the scaling exponent, \(\nu\) change with the introduction of two- and three-body DM interactions; more precisely \(h^*\) increases relative to the Heisenberg model, as verified by both quantifiers, \(\overline{\langle r\rangle}\) and \(\overline{\langle\mathcal{G}\rangle}\), as shown in Table \ref{tab:scaling_results}. On the other hand, \(h^*\) is higher in the case of a three-body DM interaction than in the case of a two-body DM interaction, indicating an increase of the ergodic region in the presence of a three-body DM interaction, as shown in Table \ref{tab:scaling_results}.  

We note that the critical disorder strengths obtained from $\overline{\langle r\rangle}$ and $\overline{\langle\mathcal{G}_1\rangle}$ do not coincide exactly, and the difference becomes more pronounced in the presence of DM
interactions, particularly for the three-body term. This discrepancy does not originate from the single-site approximation, since $\mathcal{G}_1$ approaches $\mathcal{G}$ rapidly with increasing system size (see Fig. \ref{fig:ggm_vs_gm}). Rather, the two quantities probe different features of the finite-size crossover:  the gap ratio detects the change in spectral statistics, while the GGM captures the suppression of genuine
multipartite entanglement in the eigenstates. The effective next-nearest-neighbor hopping generated by the three-body DM interaction enhances delocalization and broadens the crossover regime, thereby allowing the two indicators to acquire distinct critical disorder strengths. The values reported in Table~\ref{tab:scaling_results} should therefore be regarded as observable-dependent finite-size estimates of the critical disorder strengths.

\section{Capturing ergodic to MBL transition via dynamical quantity}
\label{sec:dynamics}

The entire analysis performed in the previous sections is based on the properties of the eigenstates and eigenspectrum of the Hamiltonian. The multisite entanglement turns out to be a beneficial quantity for determining the transition from the ergodic to the MBL phase, apart from revealing the potentiality in quantum information tasks in the latter phase. Let us now assess whether the nature of multipartite entanglement over time can also carry signatures of the phases. 

To answer this, let us consider the evolution of an initially unentangled state under a Hamiltonian residing either in the localized or in the ergodic phase. In the MBL phase, the system typically preserves substantial memory of the initial configuration owing to the inhibited transport of energy and other local observables. However, what distinguishes the MBL phase from its non-interacting counterpart, namely the Anderson localized phase, is the nontrivial manner in which quantum information, for example, entanglement, disperses across the system. A hallmark signature of this phenomenon is the logarithmic growth of the half-chain von Neumann entropy when a product state undergoes unitary dynamics governed by an MBL Hamiltonian \cite{Luitz2016}. This slow, yet steady, entanglement growth is widely recognized as a defining trait of interacting disordered systems, where strong correlations among numerous local degrees of freedom, such as spin orientations, are crucial. Notably, this behavior is absent in non-interacting settings, thereby serving as a key diagnostic of many-body effects.

At this point, it is important to emphasize that the half-block entropy captures entanglement solely across a single bipartition, thereby overlooking the genuinely multipartite character of quantum correlations that may emerge during the dynamics. To overcome this limitation and to gain a better understanding, we investigate the evolution of multipartite entanglement in terms of GGM, following the methodology established in the preceding section. 

 We consider the Néel state, \( |\psi(0)\rangle = |\uparrow \downarrow \uparrow \downarrow \ldots \uparrow \downarrow \rangle \), as the initial configuration and evolve it under a disordered interacting Hamiltonian, yielding the time-evolved state, \( |\psi(t)\rangle = e^{-iH\tilde{t}}|\psi(0)\rangle \), where $H$ is defined in Eq. (\ref{eq:hamil}) and $t=J\tilde{t}$ is the dimensionless time scale. Interestingly, the temporal behavior of GME averaged over multiple disorder realizations, indicated as $\overline{\mathcal{G}(t)}$, exhibits a signature that demarcates the MBL phase from the ergodic one, in agreement with conventional diagnostics such as the imbalance~\cite{Luitz2016}. Note that $\overline{\mathcal{G}(t)}$ only contains averaging over the disorder realization, unlike in the static case.
 To carry out this study for moderately large system-sizes $(N \sim 18)$, we employ the Chebyshev polynomial expansion technique, which enables efficient simulation of real-time dynamics in disordered quantum systems (see Appendix~\ref{app:chb}).

Initially, all disorder strengths exhibit a rapid rise of $\overline{\mathcal{G}(t)}$ [as seen in Figs. \ref{fig:ggm_dyn} (a) and (b)] because of the initial spreading of local wave packets. A characteristic peak around $0 < t < 1$ is observed, which is attributed to the spreading of excitations up to the localization length and diminishes in magnitude with increasing \(h\) (cf. Ref.~\cite{coherence_dynamics, falcao_prl_2025,ergotropy_dynmics}). At moderate times $t\gtrsim1$, $\overline{\mathcal{G}(t)}$ increases polynomially with time $t$, followed by a saturating value. In the ergodic regime (e.g. $h \leq 3$), $\overline{\mathcal{G}(t)}$ reaches a high value close to $0.3$ which is less than the maximum value $0.5$, indicating a moderately high multipartite entanglement typical of thermalizing systems. It is important to emphasize that, while the von Neumann entropy in a bipartition may still exhibit logarithmic growth in the MBL regime because of interaction-induced dephasing, the genuine multipartite entanglement does not exhibit a comparable increase. Instead, it remains largely bounded, suggesting that although local entanglement spreads, global multipartite entanglement remains constrained; a distinction that underscores the subtle nature of information dynamics in the localized phases. As \(h\) increases, both the peak height and the long-time value of $\overline{\mathcal{G}(t)}$ decrease, signaling the onset of localization. In the localized phase (e.g., \(h \gg 4.0\)), the growth is substantially suppressed and the long-time value of $\overline{\mathcal{G}(t)}$ attains a low value, after the transient behavior.  

The dynamics of multipartite entanglement shows that, while the transient timescale \(t\lesssim 1\) is similar across disorder strengths, the time required to reach a plateau becomes much longer at strong disorder. The dynamical fluctuations of the finite size systems are suppressed by the quench averaging over the realizations, with fluctuations being higher for smaller system sizes. Since the saturation time also depends on the system size, we use scaling of the multipartite entanglement with system size at a fixed time to probe the finite-time ergodic-to-MBL crossover. At times well beyond the initial transient, i.e., $t>2\times10^3$ in our case, this scaling captures the disorder dependence of the dynamics. We choose \(t=10^4\), the largest time reached in our numerical calculations, to examine scaling with system-size \(N\). Since the dynamical regime for the plateau in finite-size, is sensitive to the disorder strength, especially in the MBL regime, analysis at late times allows the effects of disorder to become visible across the system sizes considered.

In the preceding section, we observed that the presence of DM interactions shifts the localization transition to higher values of the disorder strength \(h\), thereby extending the ergodic regime. We find that this trend persists in the dynamical study of multipartite entanglement as well. To quantitatively characterize the transition, we investigate the long-time behavior of $\overline{\mathcal{G}(t)}$.

\textit{Scaling of GGM with system size at long-time.} At sufficiently long times \(t = 10^4\), the multipartite entanglement, quantified via the disorder-averaged GGM, \(\overline{\mathcal{G}(10^4)}\)
serves as a sensitive probe for distinguishing between ergodic and many-body localized regimes. In the ergodic phase, \(\overline{\mathcal{G}(10^4)}\) exhibits a growth with system-size \(N\), suggestive of a sub-volume-law scaling and reflecting the delocalized nature of eigenstates. In contrast, the MBL phase is characterized by a near saturation and decline of \(\overline{\mathcal{G}(10^4)}\) with increasing \(N\), described as a sub-area-law behavior. These scaling features are encapsulated by the functional form as
\begin{equation}
    \overline{\mathcal{G}(10^4)} \sim  cN^\beta ,
\end{equation}
where \(\beta\) defines the scaling exponent [see Figs.~\ref{fig:ggm_steady_state} (a) and (b) for behavior in two- and three-body DM interactions], and the coefficient \(c\) captures disorder-dependent offsets. We interpret the transition point as the critical disorder strength \(h^*\) at which \(\beta \to 0\), signifying a loss of extensivity in multipartite entanglement, i.e., an archetypal signature of localization. 

We study the variation of \(\overline{\mathcal{G}(10^4)}\) with \(N\) for different disorder strengths and interaction parameters [see Figs. \ref{fig:ggm_steady_state} (a) and (b)]. The scaling behavior yields a monotonic decrease in \(\beta\) as disorder \(h\) increases. This dependence is summarized explicitly in Fig. \ref{fig:ggm_steady_state} (c), where the extracted \(\beta\) values are plotted as a function of \(h\) for three different choices of Dzyaloshinskii-Moriya  interaction strengths \((D, D')\). Notably, the presence of either two-body (\(D \neq 0\)) or three-body (\(D' \neq 0\)) DM terms shifts the critical point \(h^*\) to larger values, implying an enhancement of ergodic behavior because of competing interaction-induced delocalization mechanisms.
\begin{table}[h]
    \centering
    \begin{tabularx}{0.6\linewidth}{
    >{\centering\arraybackslash}X 
    >{\centering\arraybackslash}X 
    >{\centering\arraybackslash}X }
        \rowcolor{gray!20} \(D\) & \(D'\) & \(h^*\) \\
        \Xhline{2\arrayrulewidth}
        0     & 0     & 3.4 \\
        \Xhline{2\arrayrulewidth}
        0.5   & 0     & 3.7 \\
        \Xhline{2\arrayrulewidth}
        0.0   & 0.5   & 6.4 \\
        \Xhline{2\arrayrulewidth}
    \end{tabularx}
    \caption{ Finite-size estimates of the critical disorder strength \(h^*\) for the transition estimated from the scaling of long-time entanglement [\(\overline{\mathcal{G}(10^4)}\)] for various DM interaction configurations by setting the scaling parameter $\beta \simeq 0$.}
\label{tab:dynamics_steady_transition_points}
\end{table}

In the ergodic phase, the value of $\beta \in (0,1) $ indicates a sub-volume-law scaling that suggests the injected entanglement into the steady state survives for a longer duration. 
Interestingly, deeper into the localized phase, we observe the emergence of negative values of \(\beta\), which we term sub-area-law ($\beta <0$), as shown in Fig.~\ref{fig:ggm_steady_state} (c), corresponding to a decreasing trend of \(\overline{\mathcal{G}(10^4)}\) with \(N\). This counterintuitive result signals that entanglement becomes increasingly suppressed with system size, a manifestation of the persistence of memory of the initial (product) state. Such negative scaling is most pronounced in the absence of the three-body DM interaction, indicating a more robust localization. We interpret this phenomenon as a strong signature of entanglement freezing in the deep MBL regime, where the breakdown of thermalization is so severe that even long-time dynamics fail to generate extensive correlations.

Recent experimental advances increasingly indicate that maintaining equilibrium across varying parameter regimes poses considerable challenges, particularly in strongly interacting quantum systems and the preparation of highly excited states~\cite{Sierant2025}. In contrast, studying the system's out-of-equilibrium evolution, especially under nearly unitary conditions, has emerged as a more practical and insightful approach \cite{Schreiber2015, Bloch2017} to probe the ergodic-to-MBL transition. Within this paradigm, our analysis demonstrates that the dynamics of genuine multipartite entanglement offer a powerful diagnostic of ergodicity-breaking. Specifically, the long-time value of \(\overline{\mathcal{G}(t)}\) scales with system-size as \(\overline{\mathcal{G}(10^4)} \sim N^\beta\), with the system-size exponent \(\beta\) exhibiting sharp crossover behavior near the many-body localization transition, thereby serving as a sensitive and robust indicator of the underlying phase structure (see Table~\ref{tab:dynamics_steady_transition_points} ). These findings reaffirm that multipartite entanglement dynamics captures the essential physics of the ergodic-to-MBL crossover, even in the absence of equilibrium assumptions.

\section{conclusion}

\label{sec:conclusion}

Many-body localization (MBL) emerges from the interplay between disorder and interactions in quantum many-body systems, giving rise to emergent integrability. In this work, we showed that the multipartite entanglement, specifically the generalized geometric measure (GGM), provides an information-theoretic quantity, capable of detecting the transition from the ergodic to the MBL regime, both in the equilibrium and the dynamical framework. We explored such a transition in the disordered Heisenberg spin chains under a disordered magnetic field along with the effects of two- and three-body Dzyaloshinskii–Moriya (DM) interactions on the ergodic-to-MBL transition. These asymmetric DM exchange interactions introduce new dynamics that significantly affect the localization properties of the system, indicating an expansion of the ergodic regime with the increasing DM interaction strength.  While the measure generally involves optimizations over several bipartitions, we showed that in our considered model, it was sufficient to consider only the single-site reduced states. This simplification made the GGM highly accessible, even for larger system sizes, and suitable for experimental implementation using current quantum simulators, both digital and analog alike. From an experimental perspective, the single-site approximation
of GGM may be accessed without reconstructing the complete many-body state. Since it depends only on the largest eigenvalues of the single-site reduced density matrices, these quantities can, in principle, be estimated using randomized local measurements and classical-shadow techniques~\cite{Elben_2019, Huang2020, preskill_rand_toolbox}. It is discussed in Appendix~\ref{app:randomised_measurement_toolbox} as a
possible route for experimentally detecting the multipartite entanglement employed in this work.

Empowered with such an accessible measure,  we examined the quenched average of GGM for states situated in the middle of the spectrum, within an equilibrium framework. We found that the corresponding scale-invariant disorder value robustly indicates the transition point from ergodic to localized phases. To validate this, we matched these values with standardized measures of gap ratio and long-range correlators. The results exhibit strong agreement, reinforcing the effectiveness of multipartite entanglement as a marker of the transition. A notable and universal trend emerged from our study: in the presence of Dzyaloshinskii–Moriya interactions, the transition point shifts towards stronger disorder values, extending the thermal regime. We explained such a shift as a consequence of the injection of long-range correlations into the thermal eigenstates because of the DM interaction. Such correlations are typically seen in the presence of ergodic regimes, which require stronger disorder values to induce localization in the system. 

We ventured into the non equilibrium setting to study the dynamical behavior of the multipartite entanglement while starting with the Neel state. In the transient regime, the average GGM increases polynomially with time. We investigated the behavior of the long-time average multipartite entanglement as the system-size is varied. We found that a sub-area-law-like behavior occurs when the system is in the localized phase, while in the ergodic phase, sub-volume-law scaling is observed, consistent with delocalized and thermalizing dynamics. Notably, the disorder strength at which this transition occurs in the dynamical regime closely aligns with the critical point identified through equilibrium analysis. These findings reinforce the utility of multipartite entanglement as a dynamical indicator of ergodic–MBL transitions.

\section{Data Availability Statement}
There are no publicly available research data or software supporting this manuscript. Requests for further information or data should be sent to the authors.

\acknowledgements

We thank Sudipto Singha Roy for useful discussion. We acknowledge the use of the cluster computing facility at the Harish-Chandra Research Institute. This research was supported in part by the INFOSYS scholarship for senior students. T.S, K.D.A, and A.S.D acknowledge support from the project entitled "Technology Vertical - Quantum Communication'' under the National Quantum Mission of the Department of Science and Technology (DST)  (Sanction order No. Sanction Order No. DST/QTC/NQM/QComm/2024/2 (G)). LGCL is funded by the European Union. Views and opinions expressed are however those of the author(s) only and do not necessarily reflect those of the European Union or the European Commission. Neither the European Union nor the granting authority can be held responsible for them. This project has received funding from the European Union’s Horizon Europe research and innovation programme under grant agreement No 101080086 NeQST.  This work was supported by the Provincia Autonoma di Trento, and Q@TN, the joint lab between University of Trento, FBK—Fondazione Bruno Kessler, INFN—National Institute for Nuclear Physics, and CNR—National Research Council, Italy. This research was carried out and financed within the framework of the second Swiss Contribution MAPS (Grant No.~230870).

\appendix

\section{Numerical diagonalization}
\label{app:polfed}

We use a recently developed polynomially filtered exact diagonalization (POLFED)~\cite{Sierant2020}, to obtain the eigenstates from the middle spectrum of the Hamiltonian (complex in our case) up to system-size $N=20$. This approach is based on transforming the Hamiltonian via a polynomial function, such that the problem of finding the required spectrum of the original Hamiltonian is transformed into finding the extreme eigenspectrum (via the Lanczos method~\cite{Lanczos1950, arma2019}) of the transformed Hamiltonian. Specifically, for a finite normalized Hamiltonian (see Appendix \ref{app:chb}) $\tilde{H}=\sum_{j=1}^{M} \varepsilon_j\ket{\psi_j}\bra{\psi_j}$ with eigenvalues $\varepsilon_j\in[-1,1]$, to find $n_{\varepsilon}$ eigenvectors $\ket{\psi_j}$ with eigenvalues close to $\sigma$, the function $f_\sigma(\tilde{H})$ is used, such that the set  $\{\varepsilon_j\}$ close to $\sigma$ are the extremal (minimum and/or maximum) eigenvalues of $f_\sigma(\tilde{H})$. Finally, to use the Lanczos algorithm, only $f_\sigma(\tilde{H})\ket{\Psi}$ is needed to be computed for arbitrary $\ket{\Psi}$, without computing $f_\sigma(\tilde{H})$ explicitly.

For example, $f_\sigma(\tilde{H})=(\tilde{H}-\sigma)^2$ can be used, which is a polynomial function of $\tilde{H}$ which has high density of states near $\sigma \sim 0$, i.e., near the middle of the spectrum. Therefore in this case, the Lanczos algorithm faces convergence issues and is numerically unstable, as consecutive eigenvalues are very close to each other, and a gap is needed for numerical stability~\cite{Lanczos1950}. In the shift-invert approach, $f_\sigma(\tilde{H})=(\tilde{H}-\sigma I)^{-1}$, is taken, where $I$ is the identity matrix. Such a non-analytic function produces a numerically stable eigenspectrum of $f_\sigma(\tilde{H})$ (only when $\sigma\notin\{\varepsilon_j\}$) as $\{\varepsilon_j\}$ close to \(\sigma\) have a gap inverse to the density of states near $\sigma$. The idea of POLFED is to expand the Dirac delta function $\delta(x-\sigma)$ in terms of Chebyshev polynomials, i.e., $\delta(x-\sigma)=\sum\limits_{p=0}^{\infty}c_p^\sigma T_p(x)$, with  $c_p^\sigma=2^{\min(1,p)}\cos(p\cos^{-1}\sigma)$ as the coefficients. This series is truncated to a finite order $K$, which is to be chosen based on the density of states of $\tilde{H}$ near $\sigma$ such that the $n_{\varepsilon}$ eigenvectors have transformed eigenvalues in a single band. Therefore, 
\begin{equation}
    f_\sigma^K(\tilde{H})=\frac{1}{D}\sum_{p=0}^{K}c_p^\sigma T_p(\tilde{H}),
\end{equation}
is taken for a given $K$, and $D$ is the normalization constant to ensure $f_\sigma^K(\sigma)=1$, so that the maximum eigenvalue does not exceed $1$. We use each $\tilde{H}$ as a (complex) sparse matrix of Armadillo C++ library~\cite{arma2019, arma2025} and the recursion relation of Chebyshev polynomials to compute $f_\sigma^K(\tilde{H})\ket{\Psi}$ for arbitrary vector $\ket{\Psi}$. This multiplication is written as a function, which is fed to the Arnoldi method (as during the time of this work, Armadillo does not support the complex sparse matrices for symmetric Lanczos method), by tweaking the `eigs\_gen' function of Armadillo.

\emph{Choice of $K$.} As shown in Ref. ~\cite{Sierant2020}, for a given Hamiltonian $\tilde{H}$, the order $K$ can be chosen before if the form of density of states of $\tilde{H}$. The guiding principle is that the lowest eigenvalue $\theta_{\min}$ of $f_\sigma^K(\tilde{H})$ satisfies $\theta_{\min}>0.17$, as this is the value of a next maximum of $f_\sigma^K(-1\leq\varepsilon\leq1)$, the eigenvector corresponding to  $\theta_{\min}$ can be far away from $\sigma$. For a given $n_{\varepsilon}$, $\theta_{\min}$ keeps on decreasing, along with decreasing time required for diagonalization with increasing the value of $K$, as the $f_\sigma^K(\tilde{H})$ becomes sharper near $\sigma$ with increasing $K$. We keep $K$ such that $\theta_{\min}>0.2$, and obtain $K=60$ to $100$ for $N=16$, and $K=240$ to $300$ for $N=18$ for different values of disorder strength $h\in [0.5, 10.0]$ and DM interaction strengths $D$ and $D^\prime$.

Once the $n_{\varepsilon}$ eigenvalues $\{\theta_j\}$ and the corresponding eigenvectors $\{\ket{\psi_j}\}$ are obtained with $\min\{\theta_j\}>0.2$, the eigenvalues $\{E_j\}$ of the original $H$ are computed as $E_j=\bra{\psi_j}H\ket{\psi_j}$, which are used to compute the level spacing ratios. The eigenvectors $\{\ket{\psi_j}\}$ are be used to compute the properties or observables of the system probed.

\section{Chebyshev polynomial expansion}
\label{app:chb}
The Chebyshev polynomial expansion provides a numerically stable and efficient method for simulating the time evolution of quantum states governed by a Hamiltonian~\cite{Weise2006, Weise2008}, specifically for large Hilbert space dimension. Given a quantum system described by a Hamiltonian \( H \) and an initial state \( |\psi(0)\rangle \), the time-evolved state at time \( t \) is given by \( |\psi(t)\rangle = e^{-i H t} |\psi(0)\rangle \). Instead of diagonalizing the Hamiltonian or using direct Taylor expansion, the Chebyshev method expands the evolution operator in terms of Chebyshev polynomials of the first kind, \( T_n(x) \), which are defined over the interval \([-1, 1]\). To use this expansion, the Hamiltonian \( H \) must first be rescaled such that its spectrum lies within \([-1, 1]\). This is done via a linear transformation \( \tilde{H} = \frac{H - b}{a} \), where \( a = \frac{E_{\max} - E_{\min}}{2} \) and \( b = \frac{E_{\max} + E_{\min}}{2} \), with \( E_{\min} \) and \( E_{\max} \) being the minimum and maximum eigenvalues of \( H \), respectively. The series then approximates the time evolution operator
\begin{equation}
e^{-i H t} \approx e^{-i b t} \sum_{n=0}^L c_n(t) T_n(\tilde{H}),
\end{equation}
where the coefficients \( c_n(t)=2^{\min(1,n)}(-i)^n J_n(a t) \), \( J_n \) denotes the Bessel function of the first kind of order \( n \) and $L\to \infty$ is the exact representation of the unitary. The application of Chebyshev polynomials to the state vector is performed recursively using the relations: \( |v_0\rangle = |\psi(0)\rangle \), \( |v_1\rangle = \tilde{H} |v_0\rangle \), and \( |v_{n+1}\rangle = 2\tilde{H} |v_n\rangle - |v_{n-1}\rangle \). The final evolved state is then computed as \( |\psi(t)\rangle = e^{-i b t} \sum_{n=0}^L c_n(t) |v_n\rangle \). The series is truncated at finite $L\approx\max(30, 2at)$ (Appendix C of ~\cite{Sierant2019}), with atleast $30$ terms for convergence checked with $|c_n(t) \ket{v_L}|<\epsilon_m(=10^{-10})$, where $\epsilon_m$ is the error of approximation, and can be set either by machine-precision, or precision of Lanczos method while normalizing the Hamiltonian. This method is especially advantageous for large sparse Hamiltonian, as it is stable and only requires successive applications of the Hamiltonian (or its rescaled form) to vectors, without the need for full diagonalization.

\section{Experimental estimation of $\mathcal{G}_1$ using randomized measurements}
\label{app:randomised_measurement_toolbox}

The single-site approximation $\mathcal{G}_1$ is determined by the largest eigenvalue among the single-site reduced density matrices,
\begin{equation}
    \mathcal{G}_1(\ket{\Psi}) = 1-\max_i \lambda_{\max}(\rho_i),
\end{equation}
where
\begin{equation}
    \rho_i = \operatorname{Tr}_{\overline{i}}
    \bigl(\ket{\Psi}\!\bra{\Psi}\bigr)
\end{equation}
is the reduced density matrix of the $i$th spin. Consequently, an experimental estimation of $\mathcal{G}_1$ requires access only to the
single-site reduced states, rather than reconstruction of the complete many-body quantum state. 

We note that the estimation of extremal spectral properties, particularly the largest eigenvalue of all possible subsystems of a quantum state required to compute GGM, can be efficiently approximated through the recently developed randomized measurement toolbox~\cite{preskill_rand_toolbox}. This framework allows for the construction of a so-called classical shadow \(\hat{\rho}\) of the true quantum state \(\rho\) by performing randomized local unitary and projective measurements on a computational basis. Crucially, the quality of this estimate can be quantified using the operator norm, i.e., by computing a deviation between the actual and estimated states as the probability that the estimated classical shadow $\hat{\rho}$ is $\epsilon$-close to $\rho$ is given by 
   $\mathbb{P}( \|\hat{\rho} - \rho\|_\infty < \epsilon) > 1-\kappa,$
where $\kappa$ is a failure probability that decreases with the number of measurements
and depends on the dimension of the state $\rho$. Invoking Weyl's inequality \cite{Weyl1912Dec}, which bounds the deviation in eigenvalues by the operator norm difference, the estimated maximum eigenvalue \(\lambda_{\max}^{\mathrm{est}}\) and the true maximum eigenvalue \(\lambda_{\max}^{\mathrm{og}}\) obey the inequality
\(|\lambda_{\max}^{\mathrm{og}} - \lambda_{\max}^{\mathrm{est}}| < \|\hat{\rho} - \rho\|_\infty\). 
We may use this to obtain  
\begin{equation}
 \mathbb{P}( |\lambda_{\max}^{\mathrm{og}} - \lambda_{\max}^{\mathrm{est}}| < \epsilon) > 1-\kappa,
\end{equation}
thereby asserting that the classical shadow framework can provide a faithful estimation of the maximum eigenvalues without requiring full quantum state tomography~\cite{Jones2024}. This renders the approach particularly suitable for scalable entanglement diagnostics in many-body systems, where exact state reconstruction can be expensive.

\bibliography{ref}
\end{document}